%% file: paper_mod.tex
\documentclass[11pt]{article}
\usepackage{graphicx}

\newcommand{\BaBarYear}       {02}
\newcommand{\BaBarNumber}     {02}
\newcommand{\SLACPubNumber} {9185}

\newcommand{\BaBarType}     {CONF}  
\def\ups{\Upsilon (4S)}
\def\BDstarpi{${ B^{0}} \rightarrow { D^{*-} \pi^{+}}$}

\def\BDstarrho{${ B^{0}} \rightarrow { D^{*-} \rho^{+}}$}
\def\BDstarlnu{${ B^{0}} \rightarrow { D^{*-} \ell^{+} \nu }$}
\def\ifb{\rm fb^{-1}}

\input pubboard/babarsym

\long\def\inst#1{\par\nobreak\kern 4pt\nobreak
    {\it #1}\par\vskip 10pt plus 3pt minus 3pt}

\begin{document}
{\pagestyle{empty}

\begin{flushright}
\babar-\BaBarType-\BaBarYear/\BaBarNumber \\
SLAC-PUB-\SLACPubNumber \par
\end{flushright}

\par\vskip 3cm

\begin{center}
\Large \bf \boldmath
A Measurement of the Neutral $B$ Meson Lifetime using Partially Reconstructed 
\BDstarpi \ Decays 
\end{center}
\bigskip

\begin{center}
\large The \babar\ Collaboration\\
\mbox{ }\\
\today
\end{center}
\bigskip \bigskip

\begin{center}
\large \bf Abstract
\end{center}
The neutral $B$ meson lifetime has been measured with the data collected 
by the \babar\ detector at the PEP-II storage ring during the year 2000 for a total
integrated luminosity of 20.3 $\ifb$.
The \BDstarpi \ decays have been selected with a partial reconstruction method 
in which only the fast pion from the $B^0 $ decay 
and the slow pion from $D^{*-}
\rightarrow \Dzb \pi^- $ are reconstructed. 
The $B^0 $ lifetime has been
measured to be $ 1.510 \pm 0.040 \pm 0.038~ {\rm ps} $
with a sample of $6971 \pm 241$ reconstructed signal events.
\vfill
\begin{center}
Submitted to the XXXVII$^{th}$ Rencontres de Moriond on QCD and Hadronic Interactions, \\
3/16/2002---3/23/2002, Les Arcs, France 
\end{center}

\vspace{1.0cm}
\begin{center}
{\em Stanford Linear Accelerator Center, Stanford University, 
Stanford, CA 94309} \\ \vspace{0.1cm}\hrule\vspace{0.1cm}
Work supported in part by Department of Energy contract DE-AC03-76SF00515.
\end{center}

\newpage

}

\input pubboard/authors_win2002

\setcounter{footnote}{0}

\section{Introduction}

The technique of partial reconstruction of $D^{*-}$ mesons 
(charge conjugate states are always implied), in which only 
the slow pion from $D^{*-} \rightarrow \Dzb \pi^- $ is reconstructed, has been 
widely used in the past \cite{parrec} to select large samples of reconstructed $B$ mesons.
This technique provides a way to measure the combination of the CKM unitarity triangle angles $(2\beta+\gamma)$ 
with \BDstarpi \ decays \cite{physbook}. 
The present measurement has been performed as a first step towards
the goal of an analysis measuring the angle $\gamma$. 
In this respect, the reconstruction of the signal events, the rejection 
of the background, the characterization of the various background components
and finally the study of the $\Delta t $ resolution obtained in these events
are important tools for a CP analysis. All these tools are presented here in the
context of a $B^0$ lifetime measurement.

\section{The \babar\ detector and dataset}

The data used in this analysis were collected with the \babar\ detector at the PEP-II storage ring
during 2000 and correspond to an integrated luminosity of 20.3 $\ifb$ 
collected at the $\ups$ resonance and 2.6 $\ifb$ collected 40 MeV below the resonance 
for background studies (off-peak events). 

PEP-II is an energy asymmetric storage ring, with positron and electron beam 
energies of about 3.11 and 9.0 GeV. The center-of-mass frame of the $e^+e^-$ 
collision is therefore boosted along the $z$ direction in the 
laboratory frame, 
enabling decay time-dependent measurements of $B$ mesons through vertex reconstruction.

Samples of simulated $B \bar B$ and continuum events 
were analysed through the same analysis chain as the data. 
The equivalent luminosity of the generic simulated data is approximately equal to one third of the
on-resonance data, while the equivalent luminosity of a specialized simulated sample
containing  \BDstarpi \ decays followed by  $D^{*-} \rightarrow \Dzb \pi^- $
is 2.9 times larger than the on-resonance data sample.

A detailed description of the \babar\ detector and the algorithms used for the track reconstruction, particle identification and selection of $B \bar B$ events is provided elsewhere \cite{babar_nim}; a brief
summary is given here. 

Only charged particles are used for the partial reconstruction
of the signal. Particles with momentum higher than $170~\mev/c$ are reconstructed by matching hits
in the silicon vertex tracker (SVT) with track elements in the drift chamber (DCH). Since tracks with momentum below $170~\mev/c$ do not leave signals on many wires in the DCH due to the bending induced by the 
magnetic field, they are reconstructed in the SVT alone.

Electron and muon identification is used in veto mode for the selection of the fast pion. 
Electrons are identified on the basis of the energy deposited in the 
electromagnetic calorimeter (EMC), the track 
momentum and the energy loss in the DCH. Muons are selected by requiring deep penetration
in the instrumented flux return (IFR). 

The Cherenkov light emission measured in the particle identification detector (DIRC) is employed
to reject kaons from the fast pion sample. 

Neutral particles are reconstructed from clusters in the EMC that are unmatched to projected charged tracks. 
They are used only to compute event shape quantities.

\section{The partial reconstruction technique}

The \BDstarpi \ decays are reconstructed using only the tracks from the $ \pi^+ $
(fast pion) and from the $\pi^- $ (slow pion) in $D^{*-}
\rightarrow \Dzb \pi^- $. The fast pion momentum in the $\ups$ rest frame 
is required to be
in the range kinematically allowed for the decay \BDstarpi \ . 
For nominal values of the beam energies this momentum is in the range 
2.114 - 2.404 GeV/$c$. The momentum of the slow pion is  required to be greater than 
50 MeV/$c$. 

Assuming that the fast and slow pion come from the decay \BDstarpi \ followed by the two body
decay $D^{*-}
\rightarrow X \pi^- $, it is possible to compute the mass of the recoiling system $X$ 
averaged over an unmeasured angle due to the unknown direction of the \B\ momentum in the
$\ups$ rest frame. For \BDstarpi \ decays this recoil mass $M_{rec}$ peaks at the $D^0$ mass with a width
slightly less than 3 $\mev/c^2$. Assuming that the \B\ momentum lies in the plane defined by the fast and slow pions momenta in the $\ups$ rest frame, 
it is possible to compute 
the helicity angle of the pion in the $D^{*-}$ rest frame as well as the candidate \Dzb  direction.

Since the dominant source of background is continuum events, the selection procedure aims 
at reducing this contribution.  The main requirements of this selection are the following:
\begin{itemize}
\item R2, the ratio of the second to the zeroth Fox-Wolfram moment 
\cite{fox}, computed from charged particles, is required
to be less than 0.35.
\item No other tracks should be in a cone of opening angle 0.4 rad centered on the fast pion momentum
in the $\ups$  rest frame. This cut is effective against continuum events 
because in this case tracks tend to be clustered in jets.
\item A Fisher discriminant $F_D$ is computed from 15 event shape variables. Among these are 
 the scalar sum of 
the momenta in the  $\ups$  rest frame of all tracks and neutrals, in nine $20^0$ 
angular bins around the fast pion direction. The signal peaks at $F_D= -0.5$ 
while the continuum background peaks at $F_D= 0$. 
Events are required to satisfy $F_D<-0.1$. 
\item  The cosine of the soft pion helicity angle is required to be larger than 0.4 in absolute value. 
\end{itemize}

The $B^0$ decay point is determined from a vertex fit of the fast and slow pion tracks and the beam spot
position in the plane perpendicular to the beam axis (the $x$-$y$ plane). The beam spot is determined on a run-by-run 
basis using two-prong events \cite{babar_nim}. Its size in the horizontal direction is 120 $\mu$m. 
Although the beam spot size in the $y$ direction is only a few $\mu$m, a beam spot 
constraint of 30 $\mu$m is applied to account for the flight of the $B^0$ in the $y$ direction. Only events for which 
the probability of the vertex fit is greater than 0.1\% are considered further. 

The decay point of the other $B$ is determined from a subset of the remaining tracks in the event. All the
tracks with a center-of-mass angle greater than 1 rad with respect to the candidate \Dzb  direction are considered. This requirement is used to remove most of the tracks from the decay of the 
$\Dzb$ daughter of the $D^{*-}$, which would otherwise bias the reconstruction of the other /B/ vertex position. The selected tracks are then constrained to the beam spot in the $x$-$y$ plane. The track with the largest contribution to the vertex $\chi^2$, if greater than 6, is removed 
and the fit iterated until no tracks fails this requirement. Vertices composed of just one track and the beam spot
are rejected in order to reduce the number of poorly measured vertices.  Simulation shows that after all these requirements in about 85\% of signal events the other vertex has no tracks
from the $\Dzb$ decay.

Figure \ref{fig:mreclifetime}  shows the recoil mass distribution for on-resonance data events obtained by applying all the above
cuts when the fast and slow pions have opposite-sign charges. These events are 
used to measure the $B^0$ lifetime. 
Events with same-sign charges for the two pions 
are used as a background control sample (same-sign sample). Another background control sample 
is obtained
by reversing the cut on the Fisher discriminant and requiring $F_D>0$ ($ B \bar B$ depleted sample). 

The individual distributions shown in Figure~\ref{fig:mreclifetime} are obtained by fitting simultaneously to the 
on- and off-resonance data the contributions from 
\begin{itemize}
\item \BDstarpi \ events;
\item \BDstarrho \ events;
\item \BDstarlnu \ events;
\item peaking $ B \bar B$ background, where the fast and slow pions originate from the same B meson decay;
\item non-peaking $ B \bar B$ background, excluding the events from the previous four categories;
\item peaking $c \bar c$ background, where the slow pion comes from the $D^{*-}
\rightarrow \Dzb \pi^- $ decay; 
\item continuum background excluding the events from the previous category. 
\end{itemize}

All peaking backgrounds peak in the recoil mass distribution.

The shape of each contribution has been determined from the Monte Carlo simulation while all
the normalizations have been left free in the fit except those for 
\BDstarrho \ and \BDstarlnu \ events which have been fixed to the value from the Monte Carlo simulation. The  \BDstarrho \ branching fraction is taken to 
be $ (6.8 \pm 3.4)~10^{-3}$ \cite{PDG2000} giving an estimate of 
1570 events in the final sample from this decay mode. 
The fit gives $6971 \pm 241$ \BDstarpi \ events.
In the lifetime analysis to be described later, the \BDstarrho \ 
events will be included as signal in addition to the 6971 
\BDstarpi \ events. 
 Table \ref{table:compo} reports the composition of the selected sample for 
recoil mass larger than 1.86 $GeV/c^2$ (signal region sample) as determined by this fit. 

\begin{figure}[!htb]
\begin{center}
\includegraphics[height=10cm]{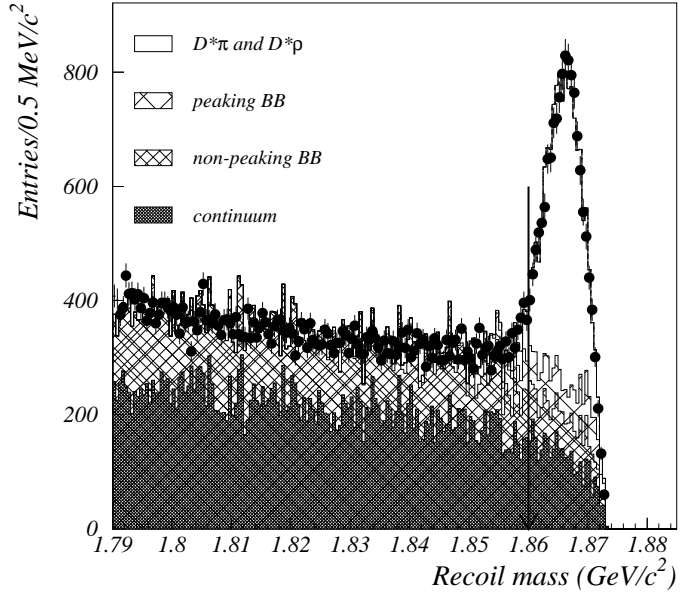}
\end{center}
\caption{
Recoil mass distribution obtained with the
selection explained in the text when the fast and slow pion have opposite-sign charges. 
In this plot the small \BDstarlnu \ component has been added to the peaking  $ B \bar B$
contribution. The continuum background includes the peaking $c \bar c$ component. 
The fit gives $6971 \pm 241 $ \BDstarpi \ events.
The region to the right of the line corresponds to the signal region. 
}
\label{fig:mreclifetime}
\end{figure}

\begin{table} [hbt]
\caption{ Composition of the data sample in the signal region. 
  }
  \centering
  \begin{tabular}{|l|c|}
    \hline
     Source &   Fraction in Signal region (\%)  \\ 
    \hline
$B^0 \rightarrow D^{*-} \pi^+ $ & 46.1   \\ 
   $B^0 \rightarrow D^{*-} \rho^+ $  & 9.3   \\
   \hline
$B^0 \rightarrow D^{*-} l^+ \nu $ & 2.3  \\
$ B \bar B$  peaking & 8.9 \\
$ B \bar B$ non peaking & 11.5   \\
$uds$ & 12.7 \\
$c \bar c$ non peak. & 5.2  \\
$c \bar c$ peaking & 4.0   \\
   \hline
  \end{tabular}
\label{table:compo}
\end{table}

\section{The lifetime measurement}

The PEP-II collider produces $B \bar B$ pairs moving along the beam direction ($z$ axis)
 with an average Lorentz boost of $\langle\beta \gamma \rangle = 0.55 $. 
The lifetime is determined by measuring the quantity 
$\Delta z = z_{decay}-z_{other}$, where $z_{decay}$ ($z_{other}$) 
is the position along the beam line of the  reconstructed 
\BDstarpi \ decay (other) vertex.
To remove badly reconstructed vertices, all events for which $\sigma_{\Delta z} > 400 \mu {\rm m}$,
where  $\sigma_{\Delta z}$ is the uncertainty on $\Delta z$ computed for each event, 
are rejected. 

Residual tracks from the $\Dzb$ decay, not removed by the track selection for the other $B$ 
vertex, bias the reconstruction of the other $B$ vertex position.
This bias is removed by correcting the $\Delta z$ value for each event with a 
correction function determined from the simulated signal sample as a
function of $\Delta z$.

The proper time difference between $B$ decays is then computed with the relation 
$\Delta t =  \Delta z / \langle c \beta \gamma \rangle$. A fit with a double Gaussian 
to the $\Delta t $ residuals in the Monte Carlo simulation shows that 
75\% of the events are contained in the narrower Gaussian, which has a width of 
 0.8 $\ps$. 

The lifetime $\tau_{B^0}$ is obtained from an unbinned maximum likelihood fit to the two-dimensional
$\Delta t$, $ \sigma_{\Delta t}$ distribution. 
The $\Delta t$ distribution of signal events is described by the convolution of the decay
probability distribution
\begin{equation}
f (\Delta t_{true} | \tau_{B^0} ) = {{ 1} \over { 2 \tau_{B^0} }} e^{-|\Delta t_{true}|/ \tau_{B^0}},
\end{equation}
with the experimental resolution function, which is represented by the sum of 
three Gaussian distributions. The first two, accounting for more than 99\% of the events,
have the form
\begin{equation}
G(\delta(\Delta t), \sigma_{\Delta t} ) = {{ 1} \over { \sqrt{2 \pi} S \sigma_{\Delta t} }}
e^{- { { (\delta(\Delta t) - b \sigma_{\Delta t} )^2   } \over { 2 S^2 \sigma_{\Delta t}^2 } } }
\end{equation}
where $\delta(\Delta t)= \Delta t-\Delta t_{true}$ is the difference between the measured and the true 
value of $\Delta t$, $b$ is a bias due to the charm tracks in the other vertex and the scale factor $S$ is introduced 
to account for possible misestimation of the error on the proper time difference
$\Delta t$. The third Gaussian of fixed bias $b=0$ and scale factor $S=6$ is added to account 
for badly mismeasured events (``outliers''). 

The \BDstarrho \ resolution function has been found with Monte Carlo simulation to be the same as for signal events. 
In the fit for the $B^0$ lifetime  the \BDstarrho \ events are considered as signal events. 
Events coming from the decay \BDstarlnu \ are added to the $ B \bar B$ peaking component and the $\Delta t$ distribution of the
latter is assumed to be equal to the $ B \bar B$ non-peaking component. 

The $\Delta t$ distribution of 
the B background contribution is described by a  resolution function similar to that used for the signal.
In this case the lifetime parameter represents an effective B lifetime which has been fitted independently from the signal lifetime.

The $\Delta t$ distribution for $uds$ events is described by the sum of three Gaussians
while the $\Delta t$ distribution for $c \bar c$ events has been described by the sum of a Gaussian
plus the same Gaussian convoluted with an exponential term to account for the effective charm hadron lifetime.
 
The function used to fit the data is the weighted sum of four contributions: 
\begin{eqnarray}
\nonumber
F(\Delta t,\sigma,\tau_{B^0}) = [1 - f_{B \bar B}(M_{rec})- f_{uds}(M_{rec})-f_{c \bar c}(M_{rec})]
F_{B^0} (\Delta t,\sigma,\tau_{B^0}) + \\
\nonumber
f_{B \bar B}(M_{rec}) F_{B \bar B} (\Delta t,\sigma) + 
f_{uds}(M_{rec}) F_{uds} (\Delta t,\sigma) + f_{c \bar c}F_{c \bar c} (\Delta t,\sigma)
\end{eqnarray}
where the functions $F_{B^0}$, $F_{ B \bar B}$, $F_{uds}$ and $F_{c \bar c}$ describe the measured decay time
difference distributions for the signal, $ B \bar B$ background, $uds$ and $c \bar c$ events, respectively.
$f_{B \bar B}$, $f_{uds}$ and $f_{c \bar c}$ are the probabilities that the event is from 
the $ B \bar B$,  $uds$ and $c \bar c$ background, computed for each event on the basis of the measured value
of the recoil mass $M_{rec}$. 

The key parameters describing the $\Delta t $ distributions for the background events 
are fitted on the $ B \bar B$ depleted control sample for the continuum background and 
on the same-sign sample for the $ B \bar B$ background. The $ B \bar B$ depleted control sample
has been used because the integrated luminosity of the off resonance sample in not sufficient 
to precisely determine the parameters of the continuum background. An alternative parameterization 
obtained by fitting the side-band control sample has also been tried giving compatible
results. 

All the parameters describing the signal resolution function are free in the fit except the bias 
of the second Gaussian which is fixed to the value found on the signal Monte Carlo sample. 

The result of the  fit to the signal region sample in the 
range $|\Delta t| < 15~\ps$ 
is $\tau_{B^0}^{raw} = 1.524 \pm 0.040 ~\ps$, where the error is statistical only. 
Figure 2 shows the comparison between the measured $\Delta t $ distribution and the fit result. 
All the fitted parameters appear in Table \ref{table:fitresult}.

\begin{figure}[htb]
\begin{center}
\mbox{
\includegraphics[height=10cm]{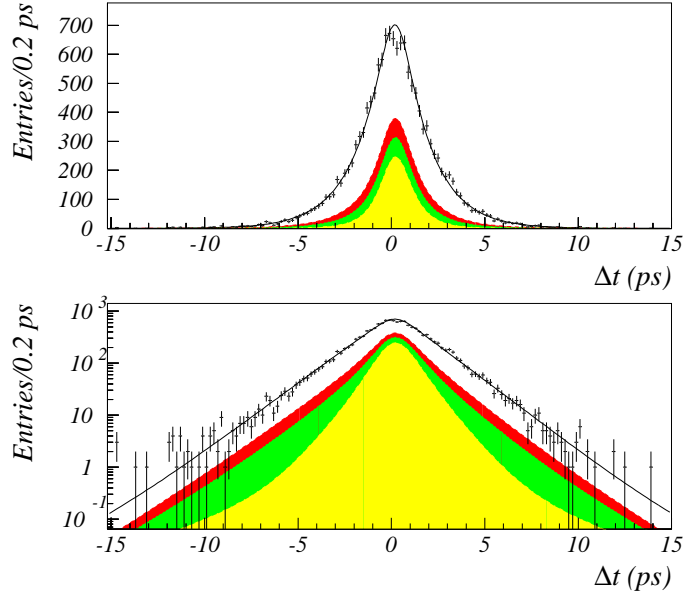}}
\end{center}
\caption{Decay time difference ($\Delta t$) distribution for the signal region sample on a linear (upper) 
and logarithmic (lower) scale. 
The curve shows the result of the unbinned maximum likelihood fit. 
The shaded areas represent from bottom to top the following contributions in the fit: 
$uds$+$c \bar c$,  non-peaking $ B \bar B$, peaking $ B \bar B$ and signal.  
}
\label{fig:fitres7}
\end{figure}

\begin{table} [hbt]
\caption{Result of the fit to the data. The $uds$ and $c \bar c$ backgrounds have been fitted
on the $ B \bar B$ depleted control sample. The $ B \bar B$  background has been fitted
on the same-charge control sample. }
  \centering
  \begin{tabular}{|l|c|c|c|}
    \hline
     Parameter description & Value \\ 
    \hline
 $uds$ narrow Gaussian fraction  & $ 0.972 \pm 0.079 $ \\
 $uds$ narrow Gaussian bias   & $0.181 \pm 0.020 $   \\ 
 $uds$ narrow Gaussian scale  & $ 1.194 \pm 0.037 $   \\ 
 $uds$ outliers fraction     & $ 0.0070 \pm 0.0028 $   \\ 
 $c \bar c$ narrow Gaussian bias    & $ 0.342 \pm 0.039 $    \\ 
 $c \bar c$ narrow Gaussian scale   & $ 1.404 \pm 0.091 $  \\ 
 $c \bar c$ lifetime fraction      & $ 0.296 \pm 0.102 $ \\ 
 $c \bar c$ lifetime  & $ 0.784 \pm 0.196 ~\ps$  \\ 
    \hline
 $ B \bar B$ lifetime  &  $ 1.611 \pm 0.044  ~\ps$   \\ 
 $ B \bar B$ narrow Gaussian fraction &  $ 0.721 \pm 0.104 $  \\ 
 $ B \bar B$ narrow Gaussian bias & $ 0.326 \pm 0.131 $  \\ 
 $ B \bar B$ narrow Gaussian scale &  $ 0.863 \pm 0.210 $  \\ 
 $ B \bar B$ outliers fraction &  $ 0.000 \pm 0.013 $  \\ 
    \hline
raw $B^0$ lifetime  & $ 1.524 \pm 0.040  ~\ps$  \\ 
signal narrow Gaussian fraction & $ 0.962 \pm 0.065 $   \\ 
signal narrow Gaussian bias &  $ 0.140\pm 0.058$   \\ 
signal narrow Gaussian scale  & $ 1.336\pm 0.082$  \\ 
signal wide  Gaussian bias & $ -0.5 $ (Fixed) \\
signal wide  Gaussian scale & $ 2.640 \pm 1.420 $  \\
signal outliers fraction  &  $ 0.0049 \pm 0.0078$   \\
    \hline
  \end{tabular}
\label{table:fitresult}
\end{table}

The raw value $\tau_{B^0}^{raw}$ value is corrected for a small bias, $0.014 \pm 0.020~\ps$,
observed when fitting the signal Monte Carlo sample using the same procedure, yielding the 
corrected value $\tau_{B^0} = 1.510 \pm 0.040~\ps$.

\section{Systematic uncertainties and cross-checks}

The systematic error on $\tau_{B^0}$ is computed by adding in quadrature the contribution from 
several sources, described below and summarised in Table \ref{table:sum_sys}.

The fractions of the various background components are varied by their uncertainties
obtained from the recoil mass fit. 
The parameters of the various background $\Delta t $ distributions are also varied by their 
uncertainties, properly accounting for their mutual correlations. 
The effective lifetime of the  $ B \bar B$ peaking component is varied by $\pm 0.044~\ps $ 
corresponding to the difference between the value found in the data and in the Monte Carlo sample. 
An alternate resolution function for this component has also been tried with negligible variation
on $\tau_{B^0}$.

The bias of the wide Gaussian, the only parameter of the signal resolution function which is not fitted, 
is varied in a conservative range. 
Several different analytical expressions are used to represent the small fraction of outliers. 
The fit range is varied from ($-10$, $10$) $\ps$ to ($-20$, $20$) $\ps$.
The parameters of the $\Delta z $ correction are  varied according to the uncertainty
due to the finite signal Monte Carlo sample size. 
The systematic uncertainty related to this correction is
estimated from the effect of a $\pm 5 \%$  variation on the fraction 
of fitted other-$B$ vertices which have no tracks from the $\Dzb$ decay.

The $z$ length scale is determined to about $0.4 \%$ from secondary interactions with a beam pipe
section of known length. 
The statistical uncertainty of the residual bias found on signal Monte Carlo events is also
added to the systematic error. 

\begin{table} [hbt]
\caption{Summary of contributions to the systematic error. }
 \centering
 \begin{tabular}{|c|c|}
    \hline
Source  & Error (fs) \\ 
    \hline
Background Parameters & 13   \\
Fractional composition & 20 \\
$\Delta z $ correction (MC) & 6  \\
$\Delta z $ correction (model) & 17  \\
Bias of the wide Gaussian (signal) & 6 \\
Outliers & 9  \\
$\Delta t$ range & 9 \\
MC bias & 20  \\
$ z$ scale  & 6 \\
    \hline
Total & 38 \\
    \hline
\end{tabular}
\label{table:sum_sys}
\end{table}

The total systematic error of $\pm 0.038 ~\ps$ is found by adding in quadrature the uncertainties from the above sources.

The dependence of the result on several different variables (angular width of the cone used to reject the
$\Dzb$ tracks, fast pion momentum, polar and azimuthal angles) has been 
studied: no statistically significant effect
is found. 

\section{Conclusion}

In conclusion the neutral $B$ meson lifetime has been measured with a sample of $6971 \pm 241$  
partially reconstructed \BDstarpi \ decays: 
$$ \tau_0 = 1.510 \pm 0.040 \pm 0.038~\ps.$$ This preliminary value is consistent with other recent \babar\
measurements \cite{bbrtaub0} and with the world average $B^0$ lifetime \cite{PDG2000}.   

\section{Acknowledgments}
\label{sec:Acknowledgments}


\input pubboard/acknowledgements

\end{document}

%% file: pubboard/authors_win2002.tex
\begin{center}
\small

The \babar\ Collaboration,
\bigskip

B.~Aubert,
D.~Boutigny,
J.-M.~Gaillard,
A.~Hicheur,
Y.~Karyotakis,
J.~P.~Lees,
P.~Robbe,
V.~Tisserand,
A.~Zghiche
\inst{Laboratoire de Physique des Particules, F-74941 Annecy-le-Vieux, France }
A.~Palano,
A.~Pompili
\inst{Universit\`a di Bari, Dipartimento di Fisica and INFN, I-70126 Bari, Italy }
% ...

\end{center}\newpage

%% file: pubboard/acknowledgements.tex
We are grateful for the 
extraordinary contributions of our \pep2\ colleagues in
achieving the excellent luminosity and machine conditions
that have made this work possible.
The success of this project also relies critically on the 
expertise and dedication of the computing organizations that 
support \babar.
The collaborating institutions wish to thank 
SLAC for its support and the kind hospitality extended to them. 
This work is supported by the
US Department of Energy
and National Science Foundation, the
Natural Sciences and Engineering Research Council (Canada),
Institute of High Energy Physics (China), the
Commissariat \`a l'Energie Atomique and
Institut National de Physique Nucl\'eaire et de Physique des Particules
(France), the
Bundesministerium f\"ur Bildung und Forschung
(Germany), the
Istituto Nazionale di Fisica Nucleare (Italy),
the Research Council of Norway, the
Ministry of Science and Technology of the Russian Federation, and the
Particle Physics and Astronomy Research Council (United Kingdom). 
Individuals have received support from 
the A. P. Sloan Foundation, 
the Research Corporation,
and the Alexander von Humboldt Foundation.